\documentclass[preprint,showpacs,preprintnumbers,amsmath,amssymb,showkeys]{revtex4}
\usepackage{graphicx}
\usepackage{dcolumn}
\usepackage{bm}

\begin{document}
\renewcommand{\thesection}{\arabic{section}}
\renewcommand{\thetable}{\arabic{table}}
\setlength{\baselineskip}{16.0pt}

\bibliographystyle{apsrev}
\title{Electron wave functions on  $T^2$ in a static \\
magnetic field of arbitrary direction}

\author{Mario Encinosa }
\affiliation{ Florida A\&M University \\Department of Physics \\
205 Jones Hall \\ Tallahassee FL 32307}
\email{encinosa@cennas.nhmfl.gov}

\begin{abstract}
A basis set  expansion is performed to find the eigenvalues and
wave functions for an electron on a toroidal surface $T^2$ subject
to a constant magnetic field in an arbitrary direction. The
evolution of several low-lying states as a function of field
strength and field orientation is reported, and a procedure to
extend the results to include two-body Coulomb matrix elements on
$T^2$ is presented.
\end{abstract}

\maketitle

\section{Introduction}

Quantum dots with novel geometries have spurred considerable
experimental and theoretical interest because of their potential
applications to nanoscience.  Ring and toroidal structures in
particular have been the focus of substantial effort
 because their topology makes it possible to explore
 Ahranov-Bohm and interesting transport phenomena
\cite{chou,datta,latge,sasaki}. Toroidal InGaAs devices have been
fabricated \cite{lorke1, garsia, mendach, zhang} and modelled,
\cite{filikhin} and toroidal carbon nanotube structures studied by
several groups \cite{sano,sasaki,shea}.

This work is concerned with the evolution of one-electron wave
functions on $T^2$ in response to a static magnetic field in an
arbitrary direction. The problem of toroidal states in a magnetic
field has been studied with various levels of mathematical
sophistication. Onofri \cite{onofri} has employed the holomorphic
gauge to study Landau levels on a torus defined by a
 strip with appropriate boundary conditions and Narnhofer has
analyzed the same in the context of Weyl algebras
\cite{narnhofer}. Here, the aim is to do the problem with standard
methodology: develop a Schrodinger equation inclusive of surface
curvature, evaluate the vector potential on that surface, and
proceed to diagonalize the resulting Hamiltonian matrix.

As noted in \cite{simonin}, ideally one would like to solve the
$N$-electron case, but the single particle problem is generally an
important first step, and while the $N$ electron system  on flat
and spherical surfaces has been studied
\cite{lorke2,bulaev,goker,bellucci,tempere,ivanov}, the torus
presents its own difficulties.  In an effort to partially address
this issue,  the evaluation of Coulombic matrix elements on $T^2$
is also discussed here.

This paper is organized as follows: in section 2 the Schrodinger
equation for an electron on a toroidal surface in the presence of
a static magnetic field is derived. In section 3 a brief
exposition on the basis set employed to generate observables is
presented. Section 4 gives results. Section 5 develops the scheme
by which this work can be extended to the two electron problem on
$T^2$, and section 6 is reserved for conclusions.
\section{Formalism}
The geometry of a
   toroidal surface of major radius $R$ and minor radius
 $a$  may be parameterized by
\begin{equation}
 \mathbf{r} (\theta,\phi)=W (\theta){\bm {\rho}} +a\  {\rm sin}
\theta{\bm  {k}}
\end{equation}
with
\begin{equation}
 W = R + a \ {\rm cos} \theta,
\end{equation}
\begin{equation}
{ \bm \rho} = \rm cos\phi {\mathbf i} + sin \phi {\mathbf j}.
\end{equation}
The differential of   Eq.(1)
\begin{equation}
d \mathbf{r}= a  d\theta \ {\bm \theta}+W d\phi{\bm \phi}
\end{equation}
with ${\bm \theta} =-\rm sin \theta  {\bm \rho}+\rm cos \theta
\mathbf{k}$ yields
  for the metric elements $g_{ij}$ on $T^2$
\begin{equation}
g_{\theta\theta}=a^2
\end{equation}
\begin{equation}
g_{\phi\phi}=W^2.
\end{equation}
The integration measure and surface gradient that follow from Eqs.
(5) and (6) become
\begin{equation}
{\sqrt g}dq^1dq^2 \rightarrow  a W d\theta d\phi
\end{equation}
and
\begin{equation}
\nabla = {\bm \theta} {1 \over a} {\partial \over \partial
\theta}+ {\bm \phi} {1 \over W} {\partial \over \partial \phi}.
\end{equation}
The  Schrodinger equation with the minimal prescription for
inclusion of a vector potential $\mathbf A$ is
\begin{equation}
 H = {1 \over {2m}}\bigg ( {\hbar
\over i} \nabla + q {\mathbf A} \bigg) ^2\Psi = E\Psi.
\end{equation}
The magnetic field under consideration will take the form
\begin{equation}
{\mathbf B} = B_1{\mathbf i} + B_0{\mathbf k},
\end{equation}
which by symmetry comprises the general case. In the Coulomb gauge
 the vector potential ${\mathbf
A}(\theta,\phi) = {1 \over 2} \mathbf{B} \times \mathbf{r} $
expressed in surface variables reduces to
\begin{equation}
\notag   \mathbf {A}(\theta,\phi) =    {1\over 2}\big [ B_1 (W
{\rm sin\phi \cos \theta}  +
 a \ {\rm sin^2\theta sin}\phi){\bm \theta}   +
 (B_0 W - B_1 a \ {\rm   sin \theta cos\phi})]{\bm \phi}
\end{equation}
\begin{equation} + B_1(F
{\rm sin\phi \sin\theta} -  a \ {\rm cos\theta sin \theta \sin
\phi})\mathbf{n}.
\end{equation}
with $\mathbf n = {\bm \phi} \ {\rm x} \  {\bm \theta}$. The
normal component of $\mathbf A$ contributes a quadratic term to
the Hamiltonian but leads to no differentiations in the coordinate
normal to the surface as per Eq.(8). There is a wealth of
literature concerning curvature effects when a particle is
constrained to a two-dimensional surface  in three-space
\cite{burgsjens,jenskoppe,dacosta1,dacosta2,matsutani,matsutani2,duclosexner,bindscatt,popov,
ouyang, midgwang,ee1,ee2, lin, goldjaffe,
exnerseba,schujaff,clarbrac}, including some dealing with the
torus specifically \cite{encmott}, but the scope of this work will
remain restricted to study of the Hamiltonian given by Eq. (9).

The Schrodinger equation (spin splitting will be neglected
throughout this work) is more simply expressed by first defining
$$\alpha = a/R$$
$$ F = 1 + \rm \alpha \ cos\theta$$
$$  \gamma_0 = B_0 \pi R^2 $$
$$ \gamma_1 = B_1 \pi R^2 $$
$$ \gamma_N = {\pi  \hbar \over q} $$
$$ \tau_0 = {\gamma_0 \over \gamma_N}$$
$$ \tau_1 = {\gamma_1 \over \gamma_N}$$
$$ \varepsilon = {2mEa^2 \over \hbar^2},$$
after which Eq. (9) may be written
$$
 \bigg [ {\partial^2 \over \partial^2 \theta} -
  {\alpha \  {\rm sin} \ \theta \over F}{\partial \over \partial
 \theta} + {\alpha^2 \over F^2}{\partial^2 \over
\partial^2 \phi} +
 i \bigg(\tau_0\alpha^2-{\tau_1\alpha^3 \over F}{\rm
sin\theta cos\phi} \bigg){\partial \over \partial \phi}
$$
$$
 + i\alpha\tau_1 {\rm sin \phi (\alpha+cos\theta)}{\partial
\over
\partial \theta}
$$
\begin{align}
 -{\tau_0^2 \alpha^2F^2 \over 4} - {\tau_1^2 \alpha^2 F^2 \over 4}
\bigg ({\rm sin^2}\phi + {\alpha^2 \ {\rm sin^2}\theta \over
F^2}\bigg) +{\tau_0 \tau_1 \alpha^3 F \over 2}\rm sin\theta
cos\phi
  \bigg] \Psi = \varepsilon\Psi
\end{align}
\begin{equation}
\Rightarrow H_\tau\ \Psi = \varepsilon \Psi.
\end{equation}

\section{Calculational scheme}
To proceed with a basis set expansion, Gram-Schmidt (GS) functions
orthogonal over the integration measure $ F = 1 + \alpha \  \rm
cos \theta$ must be generated. Fortunately, it is possible to
construct such functions almost trivially. The method for doing so
has been described elsewhere \cite{gst2}, so only the salient
results will be presented below.

The   $\tau_1 = 0, \theta \rightarrow -\theta$ invariance
 of $H_\tau$ suggests  that the solutions of the
Schrodinger equation
 be split into even and odd functions, and the primitive basis
set can be taken to possess this property;
\begin {equation} u_n(\theta) = {1 \over \sqrt
\pi} {\rm cos}[n\theta], \qquad v_n(\theta) = {1 \over \sqrt \pi}
{\rm sin}[n\theta].
\end{equation}
The  GS functions will  take the form
\begin{equation}
\psi^{\pm}_{K}(\theta) = \sum_{m}c^{\pm}_{Km} \left ( \begin{array}{c} u_m(\theta) \\
v_m(\theta) \end{array} \right)
\end{equation}
with the $c_{Km}$  given by (momentarily supressing the parity
superscripts \cite{parnote})
\begin{equation}
c_{Km}=(-)^{K+m}N_K(N_{K-1}\beta N_{K-1})(N_{K-2}\beta
N_{K-2})...(N_{m}\beta N_{m})
\end{equation}
and the  normalization factors $N_K$ determined from
\begin{equation}
N^2_{k+1}={1 \over {1-\beta^2 N^2_{k}}}
\end{equation}
starting from $N_0  = \sqrt{1/2}$ for positive parity states and
$N_1 = 1$ for negative parity states. The $K\nu^{th}$ basis state
is attained by appending   azimuthal eigenfunctions onto the GS
functions described above,
\begin{equation}
\Psi^{\pm}_{K\nu}(\theta,\phi) = {1 \over \sqrt {2
\pi}}\sum_{m}c^{\pm}_{Km} \left ( \begin{array}{c} u_m(\theta) \\
v_m(\theta) \end{array} \right)
 e^{i\nu\phi}.
\end{equation}
The matrix
\begin{equation}
H_{\tau \bar{K} K \bar{\nu} \nu}^{pq}= \big < \bar{K}^{\pm}
\bar{\nu} |H_\tau| K^{\pm} \nu \big >
\end{equation}
is then easily constructed since the matrix elements can all be
written in closed form (see the Appendix), and the eigenvalues and
eigenvectors determined here with a 30 state  expansion for each
$\theta$-parity. The ordering convention adopted for the states
was taken as
\begin{equation}
\notag \Psi^{+}_{0,-2}, \Psi^{+}_{0,-1},...\Psi^{+}_{5,2} ,
\Psi^{-}_{1,-2}...\Psi^{-}_{6,2}
\end{equation} yielding a Hamiltonian matrix blocked schematically into

\[  \left( \begin{array} {cc}
  H^{++}& H^{+-} \\
  H^{-+} & H^{--} \end{array} \right)\]
\ \
\section{Results}

Rather than present a large number of tables conveying little
useful information per unit page length, the focus will be on
indicating how some low-lying states evolve as a function of
magnetic field strength  for two distinct orientations. Some
remarks will also be made regarding the general trend seen for
higher excited states. Here the ratio $\alpha = a/R$ was set to
$1/2$ as a compromise between smaller $\alpha$ where the states
tend towards decoupled ring functions and larger $\alpha$ which
are less likely to be physically realistic.

Fig. 1 illustrates the evolution of the energy eigenvalue for five
low-lying states as a function of $\tau_0$ with $\tau_1 = 0$. The
states are all distinct and are labelled in the caption. Not shown
are values trivially obtained from the $\pm \nu B_0$ splitting
arising from $B_0 = 0, \nu \neq 0$ degeneracy. It is interesting
that level crossings with attendant movement towards a ground
state with different $K\nu$ occurs near integer values of
$\tau_0$, though it is not immediately clear if this is of real
significance. It is also of interest to show the sensitivity of
the dependence of $\Psi^*\Psi F$ on field strength. Fig. 2 shows
that even for moderate field values ($\tau_0 = 5$ corresponds to a
field of $1.3 \ T$ for a torus with $R = 50 \ nm$) the large
effective flux as compared to atomic or molecular dimensions
causes substantial modification to $\Psi^*\Psi F$ in the ground
state.

The results given in Figs. 1-2 were for a field configuration that
did not mix azimuthal basis states. To investigate an asymmetric
case, let $\tau_0 = 0$ and vary $\tau_1$ wherein no field threads
the torus. Fig. 3 is analogous to Fig. 1 as described above with
the notable exception that the $\nu$ splitting is non-trivial
(hence fewer distinct states are shown), and level crossings occur
at larger values.  There is no plot analogous to Fig. 2 for larger
values of $\tau_1$ with $\tau_0 = 0$ because $\phi$ dependence
increases rapidly in the eigenstates even for small $\tau_1$; a
contour plot is preferable. Figs. 4 and 5 show contour plot
results for two states at three field strengths. Note that there
is slightly more
 dependence in $\theta$  when $\tau_1 = 0$ for the state displayed in
Fig. 6 than in Fig. 5; the integration measure acts to cancel the
angular variation of the state displayed in Fig. 5.

\section{Extension to Coulomb integrals}
The two-electron problem on $T^2$ is complicated by the inability
(at least by the author) to find a transformation that decouples
the relative electron motion from their center of mass motion as
is easily done on $R^2$ \cite{qdhe}. The obvious transformations
do not lead to any advantage over the method adopted by workers
long used in atomic and molecular physics, which is to evaluate
the two-body matrix elements (supressing spin indices and physical
constants)
\begin{equation}\int \int \Phi^*_i({\bf r}_1)\Phi^*_j({\bf
r}_2)V({\bf r}_1,{\bf r}_2)(1-P_{12})\Phi_k({\bf r}_1)\Phi_l({\bf
r}_2)d^3{\bf r}_1d^3{\bf r}_2
\end{equation}
with
\begin{equation}
V({\bf r}_1,{\bf r}_2) = 4\pi\sum_{L,M}{1\over 2L+1} {r^l_{<}
\over r^{l+1}_{>}}
Y^*_{LM}(\theta_1,\phi_1)Y^*_{LM}(\theta_2,\phi_2).
\end{equation}  Eq. (20) can be adopted on $T^2$ subject to some
peculiarities which are due to the restriction of ${\bf r}_1,{\bf
r}_2$ to a surface.  Eq. (20) on $T^2$  with the notation employed
in section 3 becomes
\begin{equation}
\notag
 \int_0^{2\pi}... \int_0^{2\pi}
\Psi^*_{P{\nu_1}}(\theta_1,\phi_1)\Psi^*_{Q{\nu_2}}(\theta_2,\phi_2)V({\bf
r}_1,{\bf r}_2)(1-P_{12})
\end{equation}
\begin{equation}
\Psi_{R{\nu_3}}(\theta_1,\phi_1)\Psi_{S{\nu_4}}(\theta_2,\phi_2)F(\theta_1)F(\theta_2)d\theta_1
 d\theta_2 d\phi_1 d\phi_2.
\end{equation}
Consider the direct term; in terms of a spherical coordinate
system centered at the middle of the torus $( \rm r \  sin
\theta_s
 cos\phi,r \ \rm sin \theta_s
 sin\phi,r \ \rm cos \theta_s)$
\begin{equation}
{\rm r \   sin\theta_s} = R + a  \ \rm cos\theta
\end{equation}
\begin{equation}
{\rm r \ cos \theta_s} = a \rm \ sin\theta
\end{equation}
\begin{equation}
{\rm r } = \sqrt{R^2 + a^2 + 2 a R \ \rm cos\theta}
\end{equation}
and defining
\begin{equation}
\rho_{IJ} \equiv \psi^*_I(\theta)\psi_J(\theta)
\end{equation}
gives for Eq. (22) after some manipulation \begin{equation}
 \notag
\sum_{LM}\big(\delta_{M-\nu_1+\nu_3}\delta_{M+\nu_2-\nu_4}\big){(L-M)!
\over (L+M)!}\int_0^{2\pi}\int_0^{2\pi}
\rho_{PQ}(\theta_1)\rho_{RS}(\theta_2)
\end{equation}
\begin{equation}
P_{LM}(\theta_{s_1}(\theta_1))P_{LM}(\theta_{s_2}(\theta_2))F(\theta_1)F(\theta_2)
{(R^2+a^2+2a R {\rm cos\theta_<})^{L/2} \over (R^2+a^2+2a R {\rm
cos\theta_>})^{(L+1)/2}} d\theta_1d\theta_2.
\end{equation}
The arguments of the $P_{LM}$ are   evaluated  with
\begin{equation}
\theta_{s_i}= \arctan \bigg ( {R + a \ {\rm cos}\theta_i \over a \
{\rm sin} \theta_i } \bigg ).
\end{equation}
To evaluate the integral care must be taken with the $>, <$
character of the radial factor in the integrand. One way to
proceed is as follows:

1. Fix $\theta_1 = 0$. At this point $r_1$ is at its maximum;
integrate the integrand of Eq.(27) numerically over $d\theta_2$
from $[0,2\pi]$ by some suitable method to attain a value labelled
by, say, $G_0(\theta_1)$.

2. Set $\theta_1 = \delta$. Integrate $d\theta_2$ with $r_> =
r_1$, $r_< = r_2$ over the interval $[\delta, 2\pi-\delta]$, then
set $r_> = r_2$ and $r_< = r_1$ from $[2\pi-\delta, \delta]$. This
is $G_\delta(\theta_1)$.

3. Repeat the second step until the entire interval around the
toroidal cross section is covered. A table [$G_0(\theta_1),
G_\delta(\theta_1),G_{2\delta}(\theta_1)...]$  results that can
then  be integrated numerically. The exchange term proceeds
similarly; only the densities need modification.

\section{Conclusions}
 This work presents a method to  calculate  the spectrum and wave
 functions for an electron
 on $T^2$ in an arbitrary static magnetic field.
 Aside from the character of the solutions and numerical data, perhaps
the main result of this paper has to do with the ease with which
an arbitrarily large number of GS states can be  trivially
generated. Because every physical interaction can eventually be
expressed as a periodic function on $T^2$, matrix elements for the
interaction may then be evaluatued in closed form; hence the only
restriction to doing any problem on $T^2$ is matrix inversion.

The procedure employed here to generate observables lends itself
to easy incorporation of an arbitrary number of surface delta
functions or other type of potential. This is important because on
a closed nanotube, in contrast to a macroscopic crystal, there are
a relatively small number charge carriers, so the continuum
approximation may break down for smaller torii. Clearly, the
magnetic field treated here will not be sufficient to comprise the
general case as soon as any sort of lattice structure breaking
azimuthal symmetry is imposed on the torus. However, the extension
is simple to implement in Eq. (12), requiring only a few more
terms. It would be also be interesting to see the extension of the
static case discussed here to a time dependent laser control
problem of the type in \cite{pershin}.

Some remarks should be made regarding the curvature potential
$V_C$  well known to workers in the field of quantum mechanics on
curved surfaces. It was shown in \cite{eemscripta} that a full
three dimensional treatment of the problem of a particle near, but
not necessarily restricted to $T^2$, yields a spectra consistent
with inclusion of $V_C$ added to the two dimensional surface
Hamiltonian. Here the potential could not be  included without
substantially increasing the scope and complexity of the problem
undertaken. It was shown in \cite{enconeal} that the inclusion of
a vector potential precludes a separation of variables into
surface and normal degrees of freedom;  $\mathbf A$ added to the
Schrodinger equation requires solving coupled differential
equations in the surface and normal variables, or if a basis set
expansion is employed, a much more complicated procedure to
generate three-dimensional GS states.

\begin{center} {\bf Acknowledgments} \end{center} The author would like to thank B.
Etemadi for useful discussions and F. Sales-Mayor for a careful
reading of the text.

\newpage
\begin{center}{\bf Appendix}\end{center}

This appendix gives closed form expressions for the matrix
elements needed to construct the matrix $H_{\tau \bar{K} K
\bar{\nu} \nu}^{pq}= \big < \bar{K}^{\pm} \bar{\nu} |H_\tau|
K^{\pm} \nu \big >$. First let
$$
P_1 = 1 + {3 \over 2}\alpha^2
$$
$$
P_2 = 3\alpha + + {3 \over 4}\alpha^3
$$
$$
P_3 = {3 \over 2} \alpha^2
$$
$$
P_4 = {\alpha^3 \over 4}
$$
$$
f(\alpha) = {{\sqrt{1-\alpha^2}-1} \over \alpha}
$$
and define
$$
\delta_{J,K} \equiv \Delta_{J-K}
$$
 Each  operator in Eq. (9) will connect either only
like parity states or opposite parity states; no single operator
will do both. The matrix elements that connect like positive
parities are

$$
\bigg < \bar{K}^+ \bar{\nu}  \bigg | {\partial^2 \over \partial^2
\theta} \bigg | K^+ \nu \bigg > =\pi
\sum_{m=0}^{\bar{K}}\sum_{n=0}^{K} c_{\bar{K}m} c_{Kn} (-n^2)
\Delta_{\bar{\nu}-\nu}\big[ (\Delta_{m+n}-\Delta_{m-n})  +
$$
$$
{\alpha \over 2}
(\Delta_{m+n+1}+\Delta_{m-n+1}+\Delta_{m+n-1}+\Delta_{m-n-1})\big]
\eqno(A1)
$$

$$
\bigg < \bar{K}^+ \bar{\nu}  \bigg | -{\alpha \over F} \ {\rm sin}
\theta {\partial \over
\partial  \theta} \bigg | K^+ \nu \bigg > = {\alpha \pi \over 2}
\sum_{m=0}^{\bar{K}}\sum_{n=0}^{K} c_{\bar{K}m} c_{Kn}\  n \
\Delta_{\bar{\nu}-\nu}
(\Delta_{m+n-1}+\Delta_{m-n+1}-\Delta_{m-n-1}) \eqno(A2)
$$

$$
\bigg < \bar{K}^+ \bar{\nu}  \bigg | {\alpha^2 \over
F^2}{\partial^2 \over \partial^2 \phi}  \bigg | K^+ \nu \bigg > =
 {\alpha^2 \pi \over \sqrt{1-\alpha^2}}
\sum_{m=0}^{\bar{K}}\sum_{n=0}^{K} c_{\bar{K}m} c_{Kn}\ (-\nu^2) \
\Delta_{\bar{\nu}-\nu} [f^{n+m}(\alpha)+f^{|n-m|}(\alpha)]
\eqno(A3)
$$

$$
\bigg < \bar{K}^+ \bar{\nu}  \bigg | i \tau_0 \alpha^2{\partial
\over
\partial  \phi} \bigg | K^+ \nu \bigg > =\pi \tau_0 \alpha^2
\sum_{m=0}^{\bar{K}}\sum_{n=0}^{K} c_{\bar{K}m} c_{Kn} (-\nu)
\Delta_{\bar{\nu}-\nu}\big[ (\Delta_{m+n}+\Delta_{m-n})  +
$$
$$
{\alpha \over 2}
(\Delta_{m+n+1}+\Delta_{m-n+1}+\Delta_{m+n-1}+\Delta_{m-n-1})\big]
\eqno(A4)
$$


$$
\bigg < \bar{K}^+ \bar{\nu}  \bigg | -{\tau_0^2 \alpha^2 \over 4}
F^2\bigg | K^+ \nu \bigg
> =- {\pi \tau_0^2 \alpha^2 \over 4}\sum_{m=0}^{\bar{K}}\sum_{n=0}^{K}
c_{\bar{K}m} c_{Kn}  \Delta_{\bar{\nu}-\nu}
$$
$$
\big[ P_1 \Delta_{m-n}+{1\over 2}(P_2( \Delta_{m+n-1}+
\Delta_{m-n+1}+ \Delta_{m-n-1}) +
$$
$$
P_3( \Delta_{m+n-2}+ \Delta_{m-n+2}+ \Delta_{m-n-2})
$$
$$
+ P_4( \Delta_{m+n-3}+ \Delta_{m-n+3}+ \Delta_{m-n-3}) )  \big]
\eqno(A5)
$$


$$
\bigg < \bar{K}^+ \bar{\nu}  \bigg | -{\tau_1^2 \alpha^2 \over 4}
F^2 {\rm sin^2\phi} \bigg | K^+ \nu \bigg
> =- {\pi \tau_1^2 \alpha^2 \over 4}\sum_{m=0}^{\bar{K}}\sum_{n=0}^{K}
 c_{\bar{K}m} c_{Kn}  \big({1\over 2}\Delta_{\bar{\nu}-\nu}-
 {1\over 4} \Delta_{\nu -\bar{\nu}+2} - {1\over 4}
 \Delta_{\nu-\bar{\nu}-2}\big )
$$
$$
\big[ P_1 \Delta_{m-n}+{1\over 2}(P_2( \Delta_{m+n-1}+
\Delta_{m-n+1}+ \Delta_{m-n-1}) +
$$
$$
P_3( \Delta_{m+n-2}+ \Delta_{m-n+2}+ \Delta_{m-n-2})
$$
$$
+ P_4( \Delta_{m+n-3}+ \Delta_{m-n+3}+ \Delta_{m-n-3}) )  \big]
\eqno(A6)
$$

$$
\bigg < \bar{K}^+ \bar{\nu}  \bigg | -{\tau_1^2 \alpha^4 \over 4}
{\rm sin^2\theta} \bigg | K^+ \nu \bigg
> =- {\pi \tau_1^2 \alpha^4 \over 8}\sum_{m=0}^{\bar{K}}\sum_{n=0}^{K}
 c_{\bar{K}m} c_{Kn} \Delta_{\bar{\nu}-\nu}
\big[  \Delta_{m+n}+ \Delta_{m-n}+
$$
$$
{\alpha \over 4}( \Delta_{m+n+1}+ \Delta_{m-n+1}+
\Delta_{n-m+1}+\Delta_{1-m-n})
$$
$$
-{1\over 2}( \Delta_{m+n+2}+ \Delta_{m-n+2}+
\Delta_{n-m+2}+\Delta_{2-m-n})
$$
$$
-{\alpha\over 4}( \Delta_{m+n-3}+ \Delta_{m-n+3}+
\Delta_{m-n-3}+\Delta_{3-m-n}) ) \big]. \eqno(A7)
$$


\noindent The negative to negative terms are

$$
\bigg < \bar{K}^- \bar{\nu}  \bigg | {\partial^2 \over \partial^2
\theta} \bigg | K^- \nu \bigg > =\pi
\sum_{m=1}^{\bar{K}}\sum_{n=1}^{K} d_{\bar{K}m} d_{Kn} (-n^2)
\Delta_{\bar{\nu}-\nu}\big[ \Delta_{m-n}  +  {\alpha \over 2}
(\Delta_{m-n+1}+\Delta_{m-n-1})\big] \eqno(A8)
$$

$$
\bigg < \bar{K}^- \bar{\nu}  \bigg | -{\alpha \over F} \ {\rm sin}
\theta {\partial \over
\partial  \theta} \bigg | K^- \nu \bigg > = \alpha{\pi \over 2}
\sum_{m=1}^{\bar{K}}\sum_{n=1}^{K} d_{\bar{K}m} d_{Kn}\  n \
\Delta_{\bar{\nu}-\nu} (\Delta_{m-n+1}-\Delta_{n-m+1}) \eqno(A9)
$$

$$
\bigg < \bar{K}^- \bar{\nu}  \bigg | {\alpha^2 \over
F^2}{\partial^2 \over \partial^2 \phi}  \bigg | K^- \nu \bigg > =
 {\alpha^2 \pi \over \sqrt{1-\alpha^2}}
\sum_{m=1}^{\bar{K}}\sum_{n=1}^{K} d_{\bar{K}m} d_{Kn}\ (\nu^2) \
\Delta_{\bar{\nu}-\nu} [f^{n+m}(\alpha)-f^{|n-m|}(\alpha)]
\eqno(A10)
$$

$$
\bigg < \bar{K}^- \bar{\nu}  \bigg | i \tau_0 \alpha^2{\partial
\over
\partial  \phi} \bigg | K^- \nu \bigg > =
\pi \tau_0 \alpha^2 \sum_{m=1}^{\bar{K}}\sum_{n=1}^{K}
d_{\bar{K}m} d_{Kn} (-\nu) \Delta_{\bar{\nu}-\nu}\big[
\Delta_{m-n}  +
$$
$$
{\alpha \over 2}
(\Delta_{m+n+1}+\Delta_{m-n+1}+\Delta_{m+n-1}+\Delta_{m-n-1})\big]
\eqno(A11)
$$

$$
\bigg < \bar{K}^- \bar{\nu}  \bigg | -{\tau_0^2 \alpha^2 \over 4}
F^2\bigg | K^- \nu \bigg
> = - {\pi \tau_0^2 \alpha^2 \over 4}\sum_{m=1}^{\bar{K}}\sum_{n=1}^{K}
d_{\bar{K}m} d_{Kn}  \Delta_{\bar{\nu}-\nu}
$$
$$
\big[ P_1 \Delta_{m-n}+{1\over 2}(P_2( \Delta_{m-n-1}+
\Delta_{n-m-1}- \Delta_{1-m-n}) +
$$
$$
P_3( \Delta_{m-n-1}+ \Delta_{n-m-2}- \Delta_{2-m-n})
$$
$$
+ P_4( \Delta_{m-n-1}+ \Delta_{n-m-2}- \Delta_{3-n-m}) )  \big]
\eqno(A12)
$$

$$
\bigg < \bar{K}^- \bar{\nu}  \bigg | -{\tau_1^2 \alpha^2 \over 4}
F^2 {\rm sin^2\phi} \bigg | K^- \nu \bigg
> =- {\pi \tau_1^2 \alpha^2 \over 4}\sum_{m=1}^{\bar{K}}\sum_{n=1}^{K}
 d_{\bar{K}m}d_{Kn}  ({1\over 2}\Delta_{\bar{\nu}-\nu}-
 {1\over 4} \Delta_{\nu -\bar{\nu}+2} - {1\over 4}
 \Delta_{\nu-\bar{\nu}-2})
$$
$$
\big[ P_1 \Delta_{m-n}+{1\over 2}(P_2( \Delta_{m-n-1}+
\Delta_{n-m-1}- \Delta_{1-m-n}) +
$$
$$
P_3( \Delta_{m-n-2}+ \Delta_{n-m-2}- \Delta_{2-m-n})
$$
$$
+ P_4( \Delta_{m-n-3}- \Delta_{n-m-3}+ \Delta_{3-m-n}) )  \big]
\eqno(A13)
$$

$$
\bigg < \bar{K}^- \bar{\nu}  \bigg | -{\tau_1^2 \alpha^4 \over 4}
{\rm sin^2\theta} \bigg | K^- \nu \bigg
> =- {\pi \tau_1^2 \alpha^4 \over 8}\sum_{m=1}^{\bar{K}}\sum_{n=1}^{K}
 d_{\bar{K}m} d_{Kn} \Delta_{\bar{\nu}-\nu}
\big[ \Delta_{m-n}+
$$
$$
{\alpha \over 4}(  \Delta_{m-n+1}+ \Delta_{n-m+1}-\Delta_{1-m-n})
+{1\over 2}( - \Delta_{m-n+2}- \Delta_{m-n-2}+\Delta_{2-m-n})
$$
$$
+{\alpha\over 4}( - \Delta_{m-n+3}- \Delta_{m-n-3}+\Delta_{3-m-n})
) \big]. \eqno(A14)
$$

\noindent The matrix elements connecting different parities are

$$
\bigg < \bar{K}^+ \bar{\nu}  \bigg | -i {\tau_1 \alpha^3 \over F}
{\rm sin\theta cos\phi} {\partial \over \partial \phi} \bigg | K^-
\nu \bigg
> = {\pi \tau_1\alpha^3 \over 4}\sum_{m=0}^{\bar{K}}\sum_{n=1}^{K}
 c_{\bar{K}m} d_{Kn}(-\nu)
 (\Delta_{\bar{\nu}-\nu+1}+\Delta_{\bar{\nu}-\nu-1})
$$
$$
\big[\Delta_{m+n+1}+ \Delta_{n-m+1}
-\Delta_{m-n+1}-\Delta_{m+n1-1}\big] \eqno(A15)
$$

$$
\bigg < \bar{K}^+ \bar{\nu}  \bigg |\ i \alpha \tau_1 {\rm
sin\phi}
 ({\alpha +\rm cos\theta }) {\partial \over \partial \theta}
\bigg | K^- \nu \bigg
> =  \tau_1\alpha^3 \pi \sum_{m=0}^{\bar{K}}\sum_{n=1}^{K}
 c_{\bar{K}m} d_{Kn}(n)
 (\Delta_{\nu-\bar{\nu}+1}-\Delta_{\nu-\bar{\nu}-1})
$$
$$
\big[{3\alpha \over 4}(\Delta_{m+n}+ \Delta_{m-n})+
{(1+\alpha^2)\over
4}(\Delta_{m+n+1}+\Delta_{m-n+1}+\Delta_{m+n-1}+\Delta_{m-n-1})
$$
$$
+ {\alpha \over 4}
(\Delta_{m+n+2}+\Delta_{m-n-2}+\Delta_{m+n-2}+\Delta_{m-n-2})\big]
\eqno(A16)
$$

$$
\bigg < \bar{K}^+ \bar{\nu}  \bigg |{{\tau_0 \tau_1 \alpha^3}
\over 2} \ {\rm sin\theta \rm cos\phi}
 F
\bigg | K^- \nu \bigg
> =  {\tau_0 \tau_1 \alpha^3 \pi \over 4} \sum_{m=0}^{\bar{K}}\sum_{n=1}^{K}
 n \ c_{\bar{K}m} d_{Kn}
 (\Delta_{\nu-\bar{\nu}+1}-\Delta_{\nu-\bar{\nu}-1})
$$
$$
\big[{1\over 2}(1+{\alpha^2 \over 2})(\Delta_{m+n-1}-
\Delta_{m+n+1}+ \Delta_{m-n-1}- \Delta_{m-n+1} )+
$$
$$
{\alpha \over 2}(\Delta_{m+n-2}- \Delta_{m+n+2}+ \Delta_{m-n-2}-
\Delta_{m-n+2} )
$$
$$
+ {\alpha^2 \over 8}
(\Delta_{n+m+1}-\Delta_{m+n-1}+\Delta_{m-n+1}-\Delta_{m-n-1}
$$
$$
+\Delta_{n+m-3}-\Delta_{m+n+3}+\Delta_{n-m-3}-\Delta_{n-m+3})\big].
\eqno(A17)
$$
The negative to positive elements are obtained by interchanging
all indices, or equivalently, by taking their transpose.

\newpage

\bibliography{arbBbib2}

\begin{thebibliography}{45}
\bibitem[{par()}]{parnote}
\eprint{It simplifies the discussion to let parity here refer to the
  even/oddness of the $ \theta$-basis functions; labeling the true quantum
  mechanical parity will of course depend on the $\phi \rightarrow \phi +\pi$
  transformation.}
\end{thebibliography}

\newpage
\begin{center}{\bf Figure Captions}\end{center}
\noindent Fig. 1: $\varepsilon$ as a function of $\tau_0$ for five
low-lying states. Diamonds correspond to the $|\nu K^{\pm}>$ = $|0
0^+>$ state, stars to $|-1 0^+>$, squares to $|-2 0^+>$, triangles
to $|0 1^->$ and circles to $|01^+>$. \vskip 24 pt

\noindent Fig. 2: Evolution of $\Psi^* \Psi F$ for the $|00^+>$
state given for $\tau_0 = 0$ (thin line), $\tau_0=2.5$ (medium
line) and $\tau_0=5.0$ (thickest line). \vskip 24pt


\noindent Fig. 3: $\varepsilon$ as a function of $\tau_1$ for five
low-lying states. Diamonds  correspond to the $|\nu K^{\pm}>$ =
$|0 0^+>$ state, stars/squares to $|-1 0^+>$,  and
triangles/circles to $|-2 0^+>$. \vskip 24pt

\noindent Fig. 4: Sequential evolution of the $\Psi^* \Psi F$ =
$F|00^+>$ state on $T^2$ for $\tau_0 = 0$, $\tau_0=2.5$ and
$\tau_0=5.0$. \vskip 24pt

\noindent Fig. 5: Sequential evolution of the $\Psi^* \Psi F$ =
$F|-1 0^+>$ state on $T^2$ for $\tau_1 = 0$, $\tau_1=2.5$ and
$\tau_1=5.0$. The ground state variation in $\theta$ is partially
cancelled by the integration measure.

\newpage

\begin{figure}
\centering
\includegraphics{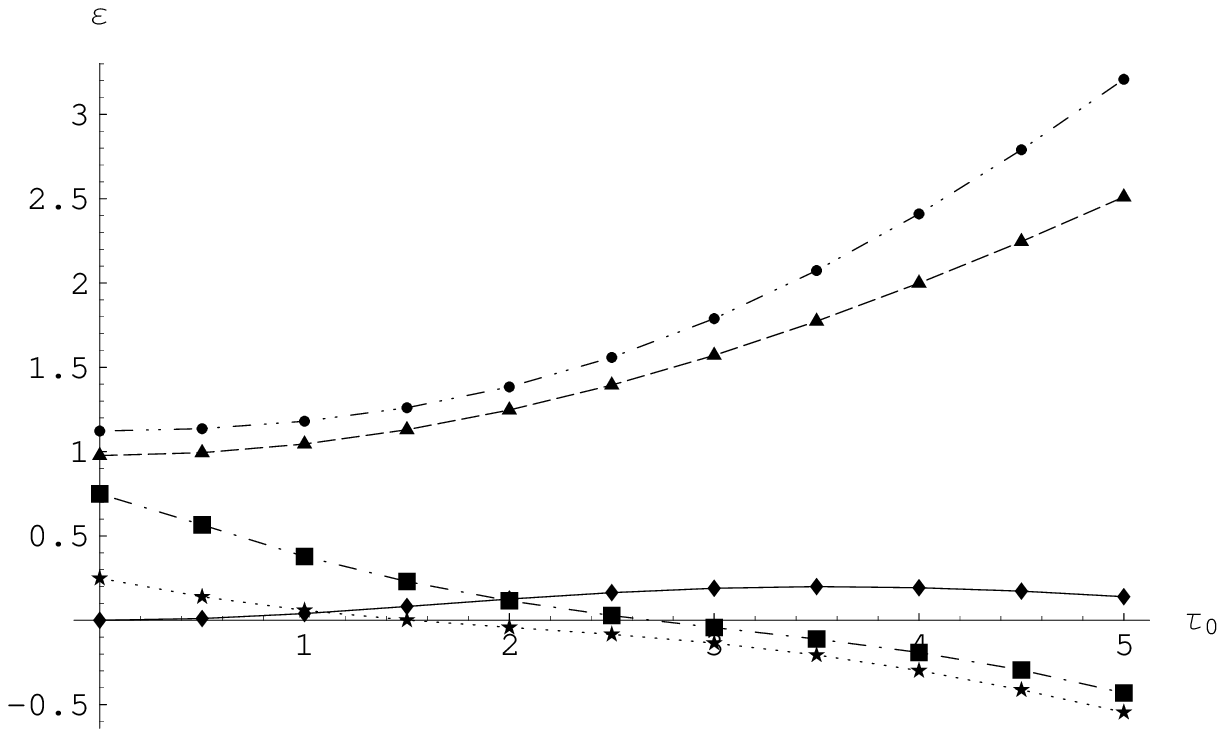}
\centerline{Fig. 1}
\end{figure}

\begin{figure}
\centering
\includegraphics{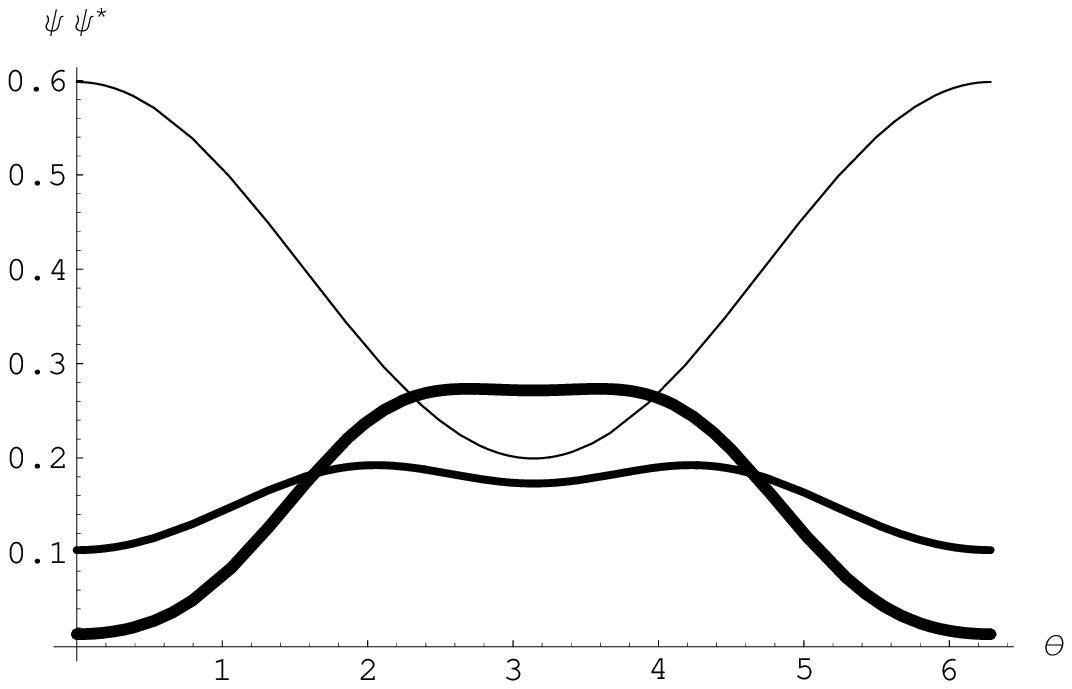}
\centerline{Fig. 2}
\end{figure}


\begin{figure}
\centering
\includegraphics{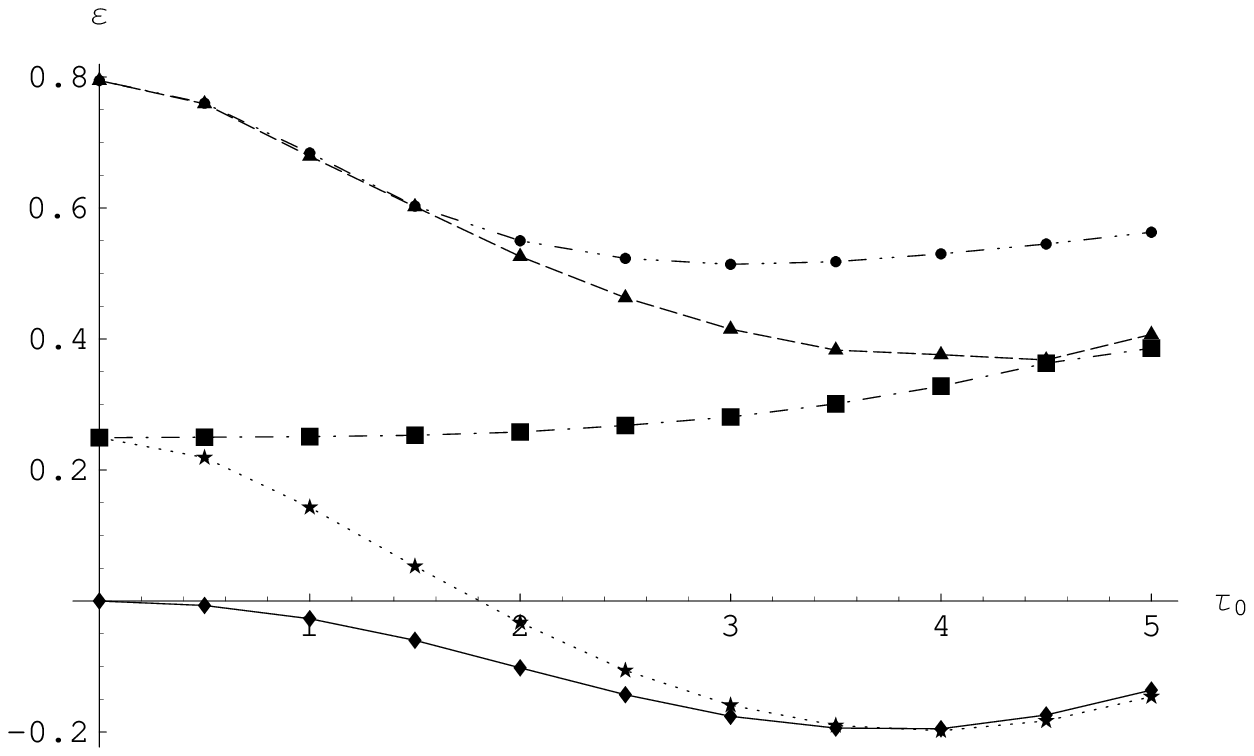}
\centerline{Fig. 3}
\end{figure}

\begin{figure}
\centering
\includegraphics{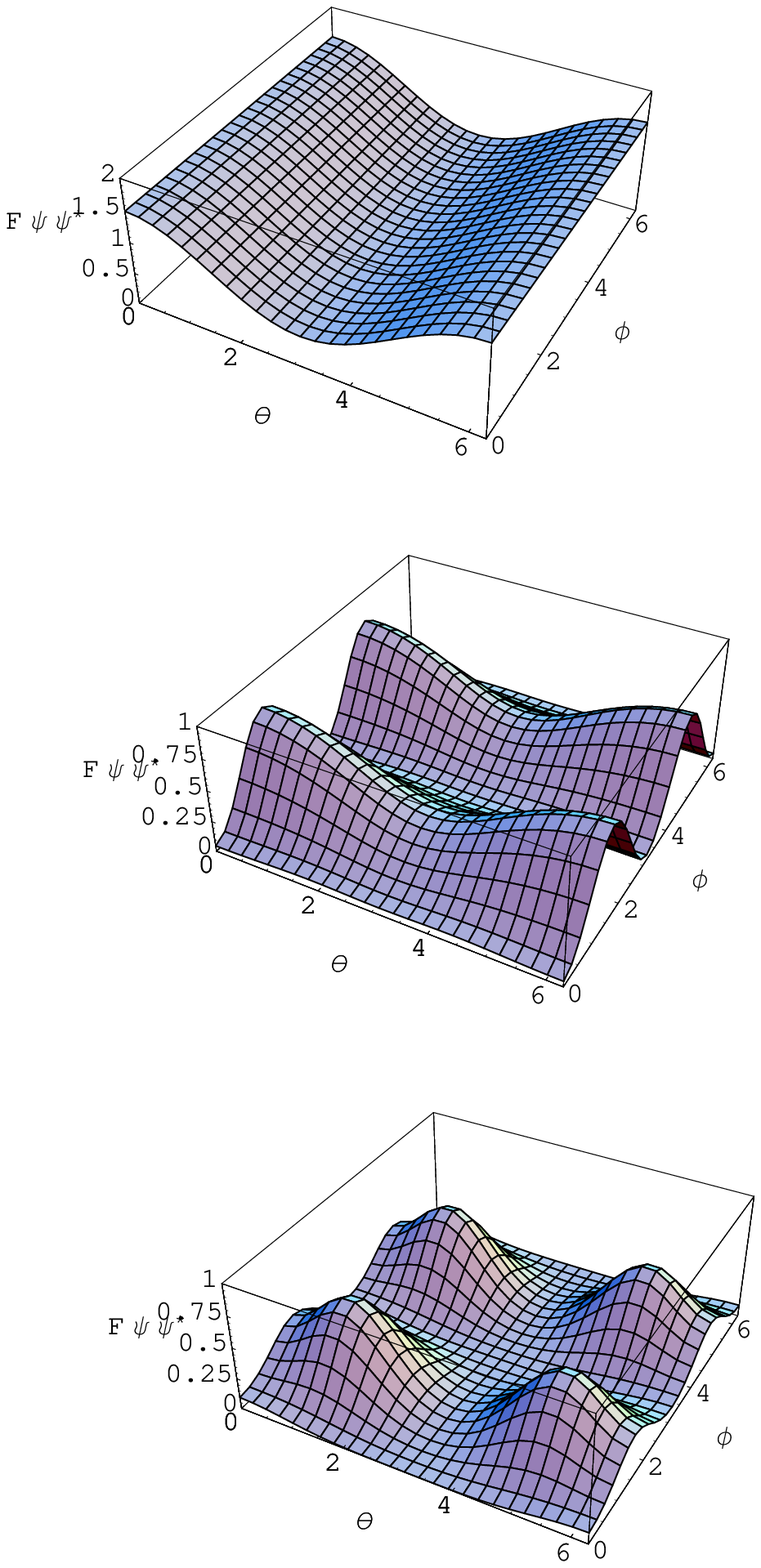}
\centerline{Fig. 4}
\end{figure}

\begin{figure}
\centering
\includegraphics{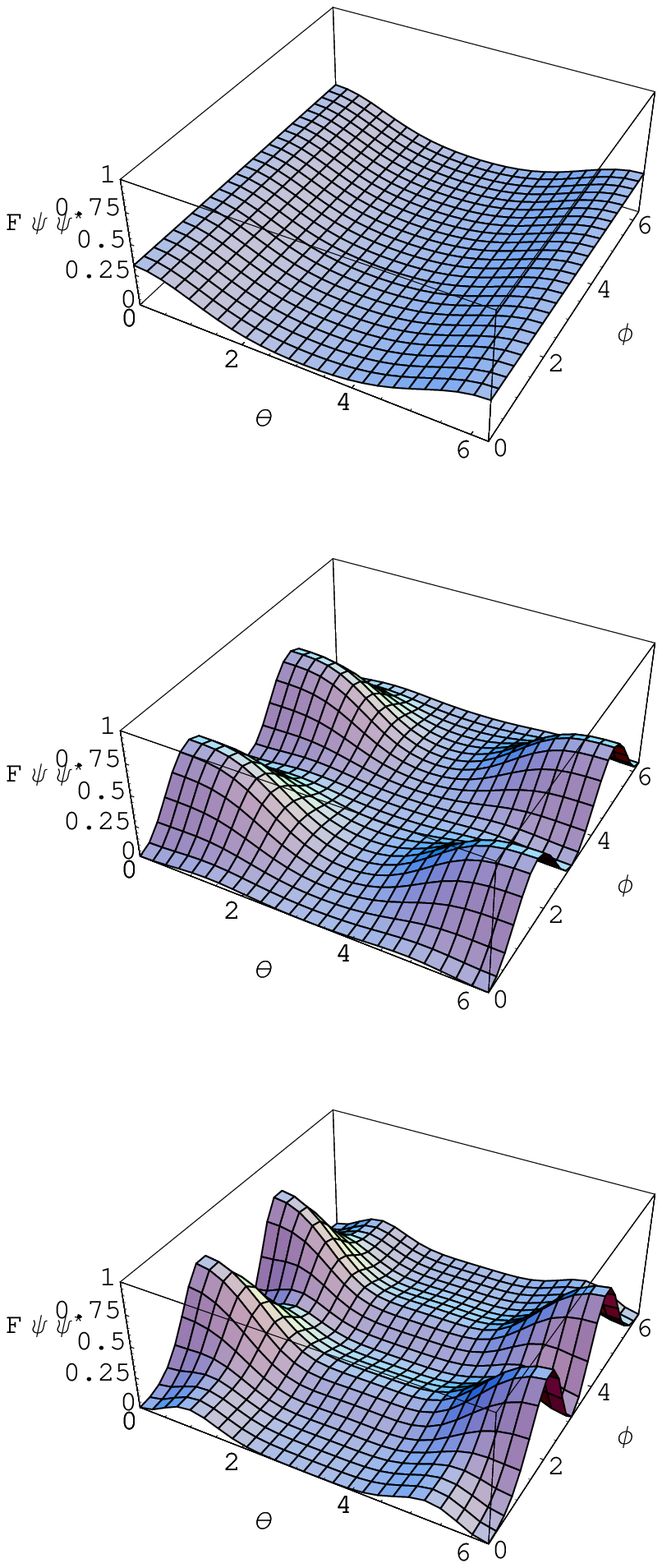}
\centerline{Fig. 5}
\end{figure}


\end{document}